\documentclass[sigconf]{acmart}
\AtBeginDocument{%
  }

\usepackage{amsmath,amsfonts}
\usepackage{algpseudocode}
\usepackage{algorithm}
\usepackage{graphicx}
\usepackage{textcomp}
\usepackage{xcolor}
\usepackage{multirow}
\usepackage{makecell}
\usepackage{balance}

\usepackage{amsthm}
\newtheorem{definition}{Definition}

\copyrightyear{2026}
\acmYear{2026}
\setcopyright{cc}
\setcctype{by}
\acmConference[WWW '26] {Proceedings of the ACM Web Conference 2026}{April 13--17, 2026}{Dubai, United Arab Emirates.}
\acmBooktitle{Proceedings of the ACM Web Conference 2026 (WWW '26), April 13--17, 2026, Dubai, United Arab Emirates}
\acmISBN{979-8-4007-2307-0/2026/04}
\acmDOI{10.1145/3774904.3792552}
\begin{document}

\title{Spattack: Subgroup Poisoning Attacks on Federated Recommender Systems}

\author{Bo Yan}
\affiliation{%
  \institution{Beijing University of Posts and Telecommunications}
  \city{Beijing}
  \country{China}
}
\email{boyan@bupt.edu.cn}

\author{Yurong Hao}
\affiliation{%
  \institution{Nanyang Technological University}
  \city{Singapore City}
  \country{Singapore}}
\email{yurong.hao@ntu.edu.sg}

\author{Dingqi Liu}
\affiliation{%
  \institution{Beijing University of Posts and Telecommunications}
  \city{Beijing}
  \country{China}
}
\email{2023213475@bupt.cn}

\author{Huabin Sun}
\affiliation{%
 \institution{Beijing University of Posts and Telecommunications}
 \city{Beijing}
 \country{China}}
\email{sunhuabin@bupt.edu.cn}

\author{Pengpeng Qiao}
\affiliation{%
  \institution{Institute of Science Tokyo}
  \city{Tokyo}
  \country{Japan}}
\email{peng2qiao@gmail.com}

\author{Wei Yang Bryan Lim}
\affiliation{%
  \institution{Nanyang Technological University}
  \city{Singapore City}
  \country{Singapore}}
\email{bryan.limwy@ntu.edu.sg}

\author{Yang Cao}
\affiliation{%
  \institution{Institute of Science Tokyo}
  \city{Tokyo}
  \country{Japan}}
\email{cao@c.titech.ac.jp}

\author{Chuan Shi}
\authornote{Corresponding author.}
\affiliation{%
  \institution{Beijing University of Posts and Telecommunications}
  \city{Beijing}
  \country{China}}
\email{shichuan@bupt.edu.cn	}

\renewcommand{\shortauthors}{Bo Yan et al.}

\begin{abstract}
Federated recommender systems (FedRec) have emerged as a promising approach to provide personalized recommendations while protecting user privacy. However, recent studies have demonstrated their vulnerability to poisoning attacks, wherein malicious clients can inject carefully crafted gradients to prompt target items to benign users. Existing attacks typically target the full user group, which compromises stealth and increases the risk of detection. In contrast, real-world adversaries may prefer to target specific user subgroup, such as promoting health supplements to older individual, to maximize attack success while preserving stealth to evade detection. Motivated by this gap, we introduce Spattack, the first poisoning attack designed to manipulate recommendations for specific user subgroups in federated setting. Specifically, Spattack adopts an approximate-and-promote paradigm, which first approximate user embeddings of target/non-target subgroups and then prompts target items to the target subgroups. We further reveal a trade-off in achieving strong attack performance on the target group while keeping the non-target group largely unaffected. To achieve a better trade-off, we propose enhanced approximation and promotion strategies. For the approximation, we push the embeddings of different subgroup away based on contrastive learning and augment the target group's relevant item set via clustering. For the promotion, we align target and relevant item embeddings to strengthen their semantic connections. An adaptive weighting strategy is further proposed to balance promotion effects between target and non-target subgroups. Experiments on three real-world datasets demonstrate that Spattack consistently achieves strong attack performance on the target subgroup with minimal impact on non-target users, even when only 0.1\% of users are malicious. Moreover, Spattack maintains competitive recommendation performance and exhibits strong resilience against mainstream defenses.
\end{abstract}

\begin{CCSXML}
<ccs2012>
   <concept>
       <concept_id>10002951.10003317.10003347.10003350</concept_id>
       <concept_desc>Information systems~Recommender systems</concept_desc>
       <concept_significance>500</concept_significance>
       </concept>
   <concept>
       <concept_id>10002978.10003006.10003013</concept_id>
       <concept_desc>Security and privacy~Distributed systems security</concept_desc>
       <concept_significance>500</concept_significance>
       </concept>
 </ccs2012>
\end{CCSXML}

\ccsdesc[500]{Information systems~Recommender systems}
\ccsdesc[500]{Security and privacy~Distributed systems security}

\keywords{federated learning, recommender systems, poisoning attack}


\maketitle

\section{Introduction}
\label{sec:introduction}
Recommender systems (RS) are essential to modern online platforms, supporting applications in e-commerce, social media, and advertising \cite{DBLP:conf/www/HeLZNHC17, DBLP:conf/sigir/Wang0WFC19, DBLP:conf/aaai/HeM16, DBLP:conf/www/SarwarKKR01}. Traditional RS rely on centralized training, which aggregates large volumes of user interaction data, raising serious privacy concerns and facing growing regulatory restrictions, such as the GDPR \cite{voigt2017eu}. 
Federated Recommendation (FedRec) has emerged as a promising solution that enables decentralized model training across user devices, avoiding directly sharing raw interaction data \cite{DBLP:journals/corr/abs-2102-04925, DBLP:journals/expert/LinLPM21, DBLP:series/lncs/YangTZCY20, DBLP:conf/aistats/McMahanMRHA17}. Typical FedRec updates user embeddings locally and uploads only item and model gradients, minimizing exposure of sensitive user-item interactions.

\begin{figure}[t]
    \centering
    \includegraphics[scale=.6]{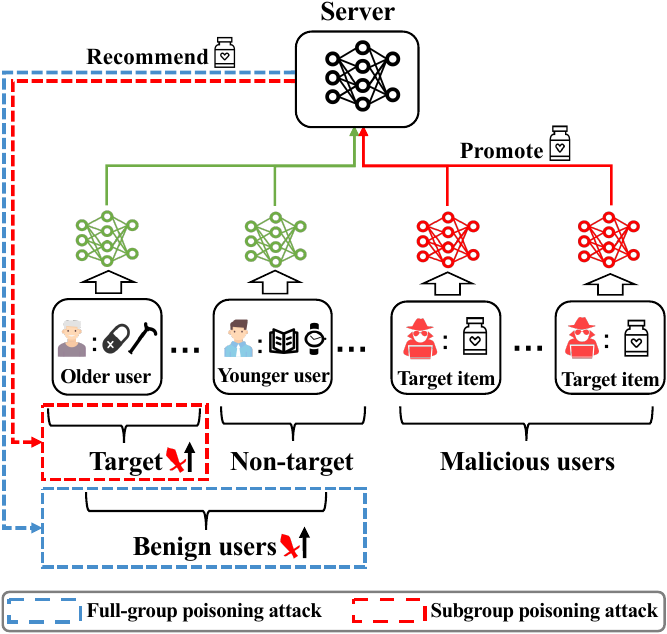}
    \setlength{\abovecaptionskip}{2pt}
    \caption{Illustration of traditional full-group poisoning attacks on all benign users and our proposed subgroup poisoning attacks on specific target users.}
    \label{fig:task}
% \vspace{-3mm}
\end{figure}

Despite these advances, recent studies have shown that FedRec remains vulnerable to poisoning attacks  \cite{DBLP:conf/wsdm/ZhangYCHNC22, DBLP:conf/icde/RongYZYCH22, DBLP:conf/aaai/00010WYYZ23}, where malicious clients might upload well-crafted model gradients to promote specific target items \cite{DBLP:conf/wsdm/ZhangYCHNC22, DBLP:conf/icde/RongYZYCH22, DBLP:conf/ijcai/RongHC22}. Poisoning attacks have attracted considerable attention due to their financial incentives and high potential for disruption. 
As illustrated in Fig.~\ref{fig:task}, the blue dashed line highlights traditional poisoning attacks, which are designed to influence all benign users by crafting updates that globally promote target items (e.g., recommending health supplements to all users). This is referred to as a \textit{full-group poisoning attack}.
% compromise the recommender systems. 
% Depending on their objective, poisoning attacks can be divided into untargeted poisoning attacks \cite{DBLP:conf/uss/FangCJG20, DBLP:conf/aaai/ZhangZWL00025} and targeted poisoning attacks \cite{DBLP:conf/wsdm/ZhangYCHNC22, DBLP:conf/icde/RongYZYCH22, DBLP:conf/ijcai/RongHC22}. Untargeted attacks focus on degrading the recommendation performance, while targeted attacks aim to promote the ranks of target items. 
% According to their objective, poisoning attacks are generally classified into untargeted attacks, which degrade overall recommendation performance \cite{DBLP:conf/uss/FangCJG20, DBLP:conf/aaai/ZhangZWL00025}, and targeted attacks, aiming to promote specific items \cite{DBLP:conf/wsdm/ZhangYCHNC22, DBLP:conf/icde/RongYZYCH22, DBLP:conf/ijcai/RongHC22}. 
Existing full-group poisoning attacks fall into two categories: (1) making target items mimic popular items \cite{DBLP:conf/wsdm/ZhangYCHNC22, DBLP:conf/www/YinXFG24, DBLP:conf/icde/Zhang0RZ0S24} and (2) approximating benign users and matching target items \cite{DBLP:conf/icde/RongYZYCH22,DBLP:conf/ijcai/RongHC22, DBLP:conf/sigir/YuanNHCY23}. The former is based on the theory of popularity bias \cite{DBLP:conf/recsys/AbdollahpouriBM17, DBLP:conf/wsdm/Zhu0ZZWC21}, where the model is prone to recommend popular items. The latter aims to approximate benign user embeddings such that directly modeling the interactions between approximated users and target items.

Although full-group poisoning attacks are verified to be effective, we argue that they overlook a more practical setting where adversaries aim to target only a specific subgroup of users, namely, \textit{subgroup poisoning attack}. As shown in Fig. \ref{fig:task}, marketers of dietary supplements have frequently deployed targeted promotions to older users (interacted with medications or canes) \cite{agny2024prevagen}. Similarly, online misinformation campaigns target vulnerable demographics such as older adults to maximize persuasion and minimize scrutiny \cite{guess2019less}. Moreover, Cambridge Analytica used Facebook to micro-target specific voter segments during the U.S. election, aiming to influence swing voters rather than the entire electorate \cite{DBLP:journals/jasss/PilditchM21}. The goal of a subgroup poisoning attack is to maximize its success rate on the target subgroup while minimizing the impact on non-target subgroup, which offers two key benefits:
(1) \textbf{Enhancing attack stealthiness}. Selective targeting not only avoids user groups that are more sensitive to unexpected recommendations (e.g., younger users \cite{DBLP:conf/ercimdl/BeelLNG13}), but also limits disruptions to the overall recommendations, thereby reducing both user- and system-level detection risks.
(2) \textbf{Improving attack effectiveness}. Some users (e.g., older users) may be more susceptible to such manipulative recommendations and targeting these users can improve the attack effectiveness.

\begin{figure}[t]
    \centering
    \includegraphics[scale=.176]{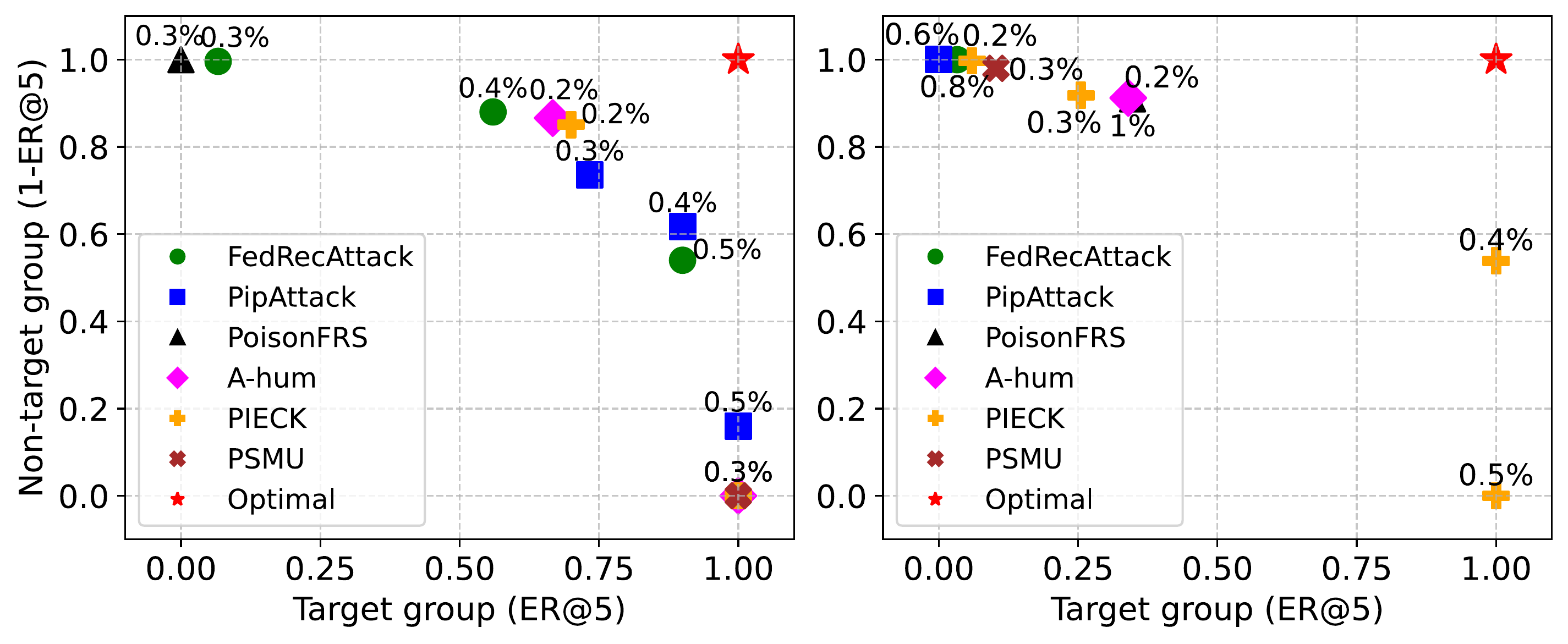}
    \caption{The trade-off dilemma of full-group poisoning attacks on target and non-target groups across various malicious client ratios (\textbf{Left}: ML-100K; \textbf{Right}: Steam). Some overlapping points are omitted for clarity, and the y-axis is set to 1-Exposure Ratio@5 (ER@5) to better visualize the trade-off.}
    \label{fig:tradeoff}
% \vspace{-3mm}
\end{figure}

Given this, a natural question arises: \textit{Can existing full-group poisoning methods be adapted to perform subgroup poisoning}? To answer this question, we conduct empirical studies on the real-world datasets. Users are divided into target and non-target subgroups based on whether they have interacted with a randomly sampled item set. Existing attacks launch attacks based on the sampled set and the performance is evaluated under varying malicious client ratios, as shown in Fig. \ref{fig:tradeoff}.
% The answer is negative. As illustrated in Fig. \ref{fig:task}, such attacks must satisfy three key criteria: (1) high attack performance on the target group, (2) low attack performance on the non-target group, and (3) high overall recommendation performance. In practice, satisfying the first two criteria simultaneously is difficult, resulting in an inherent trade-off dilemma.
% To investigate this trade-off, we conduct empirical studies on the ML-100K and Steam datasets. Users are divided into two groups based on whether they have interacted with a randomly sampled item set, and the performance of existing attacks is evaluated under varying malicious client ratios, as shown in Fig. \ref{fig:tradeoff}.
As the ratio increases, both groups' attack performances improve (moving from the top-left to the bottom-right), thus, no solution approaches the optimal point of a subgroup poisoning attack (top-right). Therefore, existing attacks struggle to achieve the goal of subgroup poisoning attacks and present a trade-off dilemma. Theoretically, most modern recommender systems typically follow similarity-based paradigms, such as collaborative filtering, which recommends items based on shared user-item interaction patterns \cite{DBLP:conf/www/SarwarKKR01, DBLP:conf/www/HeLZNHC17}. Consequently, user groups are difficult to disentangle in the representation space due to the inevitable interaction overlap, making a trade-off the only feasible solution.

In this paper, we investigate the challenge problem of subgroup poisoning attacks on FedRec and propose a novel attack model named Spattack (\underline{S}ubgroup \underline{p}oisoning \underline{attack}). Spattack builds upon an approximate-and-promote paradigm. It first approximates user embeddings of the two groups based on the attacker's interested items, and then promotes/demotes the target items to the two simulated user groups. To achieve a better trade-off between the attack performance of the two groups, we further propose approximation enhancement and promotion enhancement strategies. In the approximation process, a contrastive learning-based inter-group user embedding repulsion is designed to explicitly disentangle the representation of the two groups. To further increase the discriminability of the two groups, a clustering-based relevant item construction is then proposed to incorporate more similar items into the attack's interested items. In the promotion process, we align the constructed relevant items with the target items to improve the attack performance of the target group. Meanwhile, we propose an adaptive coefficient tuning mechanism to automatically balance the attack effects on the two groups during the federated training process.

Our major contributions are summarized as follows:

\begin{itemize}
    \item To our knowledge, we are the first to explore subgroup poisoning attacks on FedRec, a more realistic and practical setting that enhances both attack effectiveness and stealthiness.
    \item We propose Spattack, a novel subgroup poisoning attack for FedRec that maintains strong attack performance on the target group with minimal effect on non-target users. Spattack first learns discriminative group approximations to better disentangle the two user groups in the representation space, and then adaptively optimizes the promotion/demotion of target items for different groups.
    \item Extensive experiments on three real-world recommendation datasets demonstrate Spattack's superior effectiveness and stealthiness compared to existing attacks. 
\end{itemize}

% This paper follows the line of targeted poisoning attacks, along which existing work can be broadly divided into two categories: (1) making target items resemble popular items and (2) approximating benign users and matching target items. The former is based on the theory of popularity bias \cite{DBLP:conf/recsys/AbdollahpouriBM17, DBLP:conf/wsdm/Zhu0ZZWC21}, where recommenders are prone to recommend popular items to all users. Based on this property, the core challenge is to find popular items. PipAttack \cite{DBLP:conf/wsdm/ZhangYCHNC22} assumes that the popularity of items can be accessed as prior knowledge by attackers. PoisonFRS \cite{DBLP:conf/www/YinXFG24} and PIECK \cite{DBLP:conf/icde/Zhang0RZ0S24} further identify popular items based on empirical observations, such as the gradient magnitude and embedding similarity. The later aims to approximate benign user embeddings such that directly modeling the interactions between approximated users and target items. The core challenge is accurately approximating user embeddings. To achieve this, FedRecAttack \cite{DBLP:conf/icde/RongYZYCH22} holds a strong assumption that some user interactions are public available, and thus utilizes these interactions to train user embeddings. A-hum \cite{DBLP:conf/ijcai/RongHC22} and PSMU \cite{DBLP:conf/sigir/YuanNHCY23} further remove prior knowledge by randomly initializing user embeddings or approximating user embeddings with randomly selected interaction items.

\section{Preliminaries}
\label{sec:preliminary}
In this section, we first present the base FedRec framework. Then we introduce the fundamental problem settings of our attack. The summary of notations can be found in Appendix \ref{sec:app:notation}.

\subsection{Federated recommendation framework}
Following \cite{DBLP:conf/icde/RongYZYCH22, DBLP:conf/wsdm/ZhangYCHNC22}, we employ NCF \cite{DBLP:conf/www/HeLZNHC17} as the base recommender, and adopt FCF \cite{DBLP:journals/corr/abs-1901-09888} as the base FedRec framework. Nevertheless, our attack is model-agnostic and applicable to recommender models following collaborative filtering.

Let $\mathcal{U}$ and $\mathcal{V}$ denote the sets of benign users and items. Each user $u_i$ owns its local training dataset $\mathcal{D}_i$ consisting of interactions $(u_i, v_j, r_{ij})$, where $r_{ij}=1$ means $u_i$ has interacted with $v_j$. $\mathcal{V}_i^+$ and $\mathcal{V}_i^-$ denote the sets of interacted and non-interacted items for user $u_i$. Each use maintains private user embeddings $\mathbf{u}_i$ locally, while public parameters include the item embedding table $\mathbf{V}$ and other recommender model parameters $\mathbf{\Theta}$. The FedRec aims to predict scores $\hat{r}_{ij}$ for $v_j \in \mathcal{V}_i^-$ and recommends the top-$k$ ones with the highest scores. In each federated training round, a selected user $u_i$ trains public parameters and its user embedding $\mathbf{u}_i$ based on local data $\mathcal{D}_i$. The objective function is:
\begin{equation}
\mathcal{L}^{rec} = -\sum_{(u_i, v_j, r_{ij}) \in \mathcal{D}_i} r_{ij} \log \hat{r}_{ij} + (1 - r_{ij}) \log(1 - \hat{r}_{ij}).
\end{equation}
Meanwhile, $\mathbf{u}_i$ is updated locally. The gradients of $\mathbf{\Theta}$ and $\mathbf{V}$ are uploaded to the server for aggregation and updating.

% Meanwhile, the user embedding $\mathbf{u}_i$ is updated locally:
% \begin{equation}
% \mathbf{u}_i^t = \mathbf{u}_i^{t-1} - \eta \cdot \nabla \mathbf{u}_i^{t-1},
% \end{equation}
% where $\eta$ is the learning rate. The gradients of public parameters $\nabla \mathbf{\Theta}_i^{t-1}$ and $\nabla \mathbf{V}_i^{t-1}$ are uploaded to the server for aggregation and updating by:
% \begin{align}
% \mathbf{V}^t &= \mathbf{V}^{t-1} - \eta \sum_{u_i \in \mathcal{U}_{t-1}} \nabla \mathbf{V}_i^{t-1}, \\
% \mathbf{\Theta}^t &= \mathbf{\Theta}^{t-1} - \eta \sum_{u_i \in \mathcal{U}_{t-1}} \nabla \mathbf{\Theta}_i^{t-1}.
% \end{align}

\subsection{Problem settings}

Let $\mathcal{V}^{in}$ denote the interested item set of the attacker. We define the \textit{target user group} as follows:

\begin{definition}[Target User Group]
	\label{def:user_group}
Given interested items $\mathcal{V}^{in}$, if a user $u_i \in \mathcal{U}$ has interacted with $\forall v_j \in \mathcal{V}^{in}$ (i.e., $\mathcal{V}^{in}  \subset \mathcal{V}_i^+$), then the user is called a \textbf{target user}. All the target users consist of the \textbf{target user group} $\mathcal{U}^t \subset \mathcal{U}$ and the remaining users consist of the \textbf{non-target user group} $\mathcal{U}^n = \mathcal{U} \backslash \mathcal{U}^t$.
\end{definition}

% As shown in Fig. \ref{fig:task}, the target user group is identified if each user in this group has interacted medications and canes (interested items). Note that the interested items can be arbitrarily selected by attackers based on the target user group. For example, if attackers target at elder people, the interested items can be selected as medications and other relevant items.

As shown in Fig. \ref{fig:task}, a target user group (older users) is identified if each user in this group has interacted a set of interested items (medications and canes). Note that the attacker can arbitrarily choose interested items solely based on public item semantics related to the target group. This flexibility allows the proposed attack to be easily adapted for different malicious objectives. For example, the interested items for online gambling can be easily selected as betting apps, in-game credits, or other semantically related items.

\subsubsection{Attacker's knowledge}
We strictly follow the standard settings of FedRec where the attackers only possess the following knowledge: (1) the public parameters $\mathbf{\Theta}_e$ and $\mathbf{V}_e$ at each training step $e$, (2) all malicious users' local models and uploaded gradients, which also can be arbitrarily modified. Besides, the attack does not allow any changes to system settings such as the learning rate.

%and (3) the global hyperparameters of FedRec systems. 
% Apart from these, the attack assumes no prior knowledge (e.g., item popularity, public interactions, or aggregation rules) and does not allow any changes to system settings such as the learning rate.

\subsubsection{Attacker's goal}
Let $\widetilde{\mathcal{U}}$ denote the set of malicious users and $\mathcal{V}_i^{topk}$ denote the top-$K$ recommended items for user $u_i$, the \textit{Exposure Ratio at rank $K$} (ER@K) of the target items $\widetilde{\mathcal{V}}$ for a user subgroup $\mathcal{U}^s$ is defined as:
\begin{equation}
\label{eq:5}
\text{ER@K} (\mathcal{U}^s) = \frac{1}{|\widetilde{\mathcal{V}}|} \sum_{{v_j} \in \widetilde{\mathcal{V}}}  \frac{|\{u_i \in \mathcal{U}^s \mid v_j \in \mathcal{V}_i^{topk}\}|}{|\{u_i \in \mathcal{U}^s \mid (u_i,v_j, 1) \notin \mathcal{D}_i\}|}.
\end{equation}

We use ER@K ($\mathcal{U}^s$) to measure the attack performance on $\mathcal{U}^s$. Thus, the goal of subgroup poisoning attack is to simultaneously maximize the ER@K ($\mathcal{U}^t)$ on target group $\mathcal{U}^t$ and minimize the ER@K ($\mathcal{U}^n$) on non-target group $\mathcal{U}^n$. To evaluate the overall performance on two groups, we further propose a new metric called $\gamma$-\textit{Group Exposure Ratio at rank K} $(\gamma\text{-GER@K})$, which is defined as: 
\begin{equation} 
\label{eq:6}
\gamma \text{-GER@K} = \gamma \text{ER@K}(\mathcal{U}^t)+(1-\gamma) (1-\text{ER@K}(\mathcal{U}^n)),
\end{equation}
where $\gamma \in [0,1]$ is the weighting parameter to balance the importance between the target and non-target groups. In practice, $\gamma$ can be flexibly adjusted to accommodate different attack requirements.

\begin{figure}[t]
    \centering
    \includegraphics[scale=0.57]{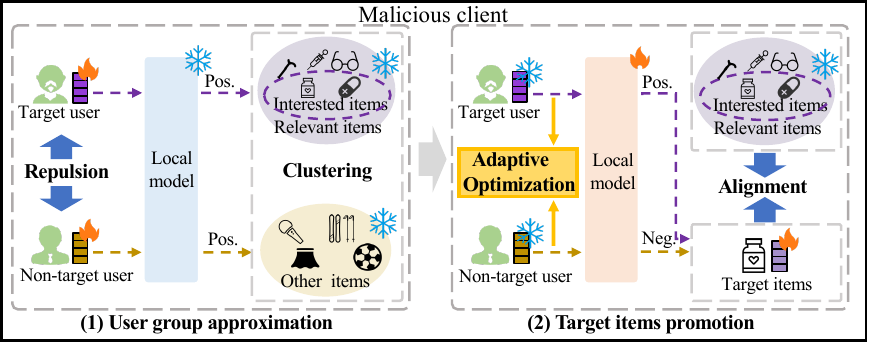}
    \setlength{\abovecaptionskip}{2pt}
    \caption{The overall framework of Spattack.}
    \label{fig:framework}
\end{figure}

\section{Methodology}
In this section, we detail our proposed Spattack. The overall framework of Spattack is depicted in Fig. \ref{fig:framework}. We first introduce the basic approximate-and-promote paradigm of Spattack, which first approximates user group embeddings using interested items and then promotes or demotes target items for different groups. Next, we present enhancement strategies for both approximation and promotion to improve the attack trade-off between target and non-target groups. We give the overall algorithm of Spattack in Appendix \ref{sec:app:alg}. 

\subsection{The basic approximate-and-promote paradigm}
One research line of existing poisoning attacks is popularity manipulation \cite{DBLP:conf/icde/Zhang0RZ0S24, DBLP:conf/wsdm/ZhangYCHNC22, DBLP:conf/www/YinXFG24}, which aligns target items with popular ones to prompt their rankings. However, this approach has fundamental limitations in subgroup poisoning settings, as popular items tend to be recommended indiscriminately to all users, resulting in the same effects for target items. Another line is user approximation \cite{DBLP:conf/icde/RongYZYCH22,DBLP:conf/ijcai/RongHC22, DBLP:conf/sigir/YuanNHCY23}, which inspires the straightforward idea that approximating two user groups, rather than traditional single one, enables more flexible manipulation of target items toward any desired group. Based on this insight, we propose the basic approximate-and-promote paradigm for subgroup poisoning attacks.

\noindent\textbf{User group approximation.} For a malicious client $\widetilde{u}_i$, we want to synthesize two kinds of embeddings $\widetilde{\mathbf{u}}_i^t$ and $\widetilde{\mathbf{u}}_i^n$ to approximate target and non-target user groups respectively. Since the attack's interested items $\mathcal{V}^{in}$ are highly relevant to the target user group (e.g., medications for the older people), we can directly utilize $\mathcal{V}^{in}$ as the user's interacted items and form the target group training set $\widetilde{\mathcal{D}}_i^t$. In this way, the target user group approximation loss is defined as: 

\begin{equation}
\label{eq:app_t}
\mathcal{L}^{app\_t}_i = -(\sum_{v_j \in \mathcal{V}^{in}} \log \hat{r}_{ij} +\sum_{v_t \in \widetilde{\mathcal{V}}}\log (1- \hat{r}_{it})).
\end{equation}
Noting that we also utilize the target item set $\widetilde{\mathcal{V}}$ as negative samples to simulate the hard users \cite{DBLP:conf/ijcai/RongHC22}, which means that target items are sampled as their negative samples.

For the non-target user group, we randomly sample an item set $\mathcal{V}^n$ with the same length as the interested item set $\mathcal{V}^{in}$ and ensure they have not appeared in $\mathcal{V}^{in}$ and target items $\widetilde{\mathcal{V}}$. Similarly, the loss function of non-target user group approximation is:
\begin{equation}
\label{eq:app_n}
 \mathcal{L}^{app\_n}_i = -\sum_{v_j \in \mathcal{V}^n} \log \hat{r}_{ij} .
\end{equation}

Combining the losses of the two groups, we can obtain the overall approximation objective:

\begin{equation}
\label{eq:app}
\arg\min_{\widetilde{\mathbf{u}}_i^t, \widetilde{\mathbf{u}}_i^n} \{\mathcal{L}^{app}_i = \mathcal{L}^{app\_t}_i + \mathcal{L}^{app\_n}_i \}.
\end{equation}
In practice, we will approximate multiple user embeddings for each group to stabilize the attack performance. In the approximation process, the only trainable parameters are user embeddings $\widetilde{\mathbf{u}}_i^t$ and $\widetilde{\mathbf{u}}_i^n$. The approximated target user embeddings are expected to resemble benign users who interacted with the interested items, while non-target embeddings resemble other users.

\noindent\textbf{Target item promotion.}
After obtaining approximated user group embeddings, the target item embeddings can be manipulated to either approach or diverge from users in the embedding space. For the target user group, we aim to promote the target items to them. Thus, we can utilize target items as positive samples to train target item embeddings and minimize the following loss function:

\begin{equation}
\label{eq:pro_t}
\mathcal{L}^{pro\_t}_i = -\sum_{ v_j \in \widetilde{\mathcal{V}}} \log \hat{r}_{ij}.
% =-\sum_{ v_j \in \widetilde{\mathcal{V}}} \log (\widetilde{\mathbf{u}}_i^t) ^{\top} \mathbf{v}_j .
\end{equation}
For the non-target user group, we aim to demote the target items to them. Thus, we let the target items as negative samples to train target item embeddings. Then the loss function is:
\begin{equation}
\label{eq:pro_n}
\mathcal{L}^{pro\_n}_i = -\sum_{ v_j \in \widetilde{\mathcal{V}} } \log (1-\hat{r}_{ij}).
% = -\sum_{ v_j \in \widetilde{\mathcal{V}} } \log (1- (\widetilde{\mathbf{u}}_i^n) ^{\top} \mathbf{v}_j) .
\end{equation}

Combining the losses depicted in Eq. (\ref{eq:pro_t}) and Eq. (\ref{eq:pro_n}), the overall promotion objective is:
\begin{equation}
\label{eq:pro}
\arg\min_{\widetilde{\mathbf{V}},\mathbf{\Theta}} \{\mathcal{L}^{pro}_i = \mathcal{L}^{pro\_t}_i+\mathcal{L}^{pro\_n}_i \}.
\end{equation}
In the promotion step, the target item embeddings $\widetilde{\mathbf{V}}$ and model parameters $\mathbf{\Theta}$ are trainable, and the approximated user group embeddings are fixed. The gradients of $\widetilde{\mathbf{V}}$ and $\mathbf{\Theta}$ by optimizing Eq. (\ref{eq:pro}) are uploaded to the server for aggregation.

\subsection{Target/non-target trade-off enhancement}
\label{sec:trade-off enhancement}
As discussed, a trade-off dilemma exists between attack effectiveness on target and non-target groups. As an extreme case, traditional full-group attacks do not differentiate between target and non-target users, leading to a poor trade-off between these two groups, as is shown in Fig. \ref{fig:tradeoff}. Ideally, the attacker knows the exact items interacted with by target users, enabling accurate approximation and promotion. However, this information is difficult to infer under current defenses \cite{lin2021fr}. Therefore, we strive for a better trade-off, i.e., maximize the attack effectiveness on the target group while minimizing the impact on the non-target group. Thus, based on the basic approximate-and-promote paradigm, we propose approximation and promotion enhancement strategies. 

\subsubsection{Approximation enhancement}
In the basic approximation, the user groups are approximated solely based on the interested items, resulting in an inaccurate characterization of real users. For example, the attacker aims to attack older people and selects some interested item about medications. However, older users may interact with more than just medications, and younger users may also interact with medications. Such item interaction entanglement makes the basic approximation lack sufficient discriminative capacity on real target and non-target user groups. To overcome this limitation, we propose two enhancement strategies, namely, inter-group user embeddings repulsion and relevant items construction.

\noindent \textbf{Inter-group user embeddings repulsion}. To improve the discriminative ability between the two user groups, we first propose a contrastive-based method to explicitly push the two group embeddings far away. Concretely, for an approximated target user embedding $\widetilde{\mathbf{u}}_i^t$ and a non-target embedding $\widetilde{\mathbf{u}}_i^n$, we maximize their distances in embedding space by minimizing the following loss:

\begin{equation}
\label{eq:distance}
 \mathcal{L}^{dis}_i = \frac{1}{2}[\max(0, \delta -\Vert \widetilde{\mathbf{u}}_i^t - \widetilde{\mathbf{u}}_i^n  \Vert_2
]^2.
\end{equation}
where $\Vert \cdot \Vert_2$ is the $L_2$ distance. $\delta$ denotes the margin that controls the extent to which the two embeddings are pushed apart. A larger $\delta$ encourages more discriminative representations between the two groups. The effects of $\delta$ will be discussed in the experiments. For multiple user approximations, we compute the average $L_2$ distance for each pair of user embeddings between the target group and the non-target groups. 

\noindent \textbf{Relevant items construction}. In practice, attackers lack prior knowledge of user groups, and thus our attack must ensure flexible selection of interested items without precise group information. From a popularity view, choosing popular items makes target and non-target groups less distinguishable, whereas selecting unpopular items makes it difficult to identify enough target users. From a quantitative perspective, too many interested items may weaken the identification of target users, while too few may result in insufficient discriminative ability. Nevertheless, attackers may prioritize ensuring a sufficient number of target users. Given this, assuming a small set of interested items, we aim to augment them to enhance the discriminative ability between the two groups. 

% The basic non-target user approximation in Eq. (\ref{eq:app_n}) constructs the positive set $\mathcal{V}^n$ by randomly sampling items other than the interested items. However, 
The randomly sampled items may be similar to the interested items in Eq. (\ref{eq:app_n}), causing the approximated non-target users to resemble the target users, thereby reducing discriminability. To address this, we propose a clustering-based interested item augmentation strategy, which expands the interested item set $\mathcal{V}^{in}$ by incorporating the top-$k$ most similar items, forming a new relevant item set $\mathcal{V}^{re}$. In this way, Eq. (\ref{eq:app_t}) uses $\mathcal{V}^{re}$ instead of $\mathcal{V}^{in}$ as positive item set and $\mathcal{V}^n$ in Eq. (\ref{eq:app_n}) should also exclude $\mathcal{V}^{re}$.

To obtain top-$k$ similar items, a naive approach involves computing pairwise distances between each interested item and all other items, resulting in a computational complexity of $\mathcal{O}(|\mathcal{V}|^2)$. Inspired by \cite{bendada2023consistency} that averaged item embeddings can approximately retain semantic patterns, we first compute the averaged embedding of the interested items and evaluate the cosine similarity between each item and this average embedding. Then we select the top-$k$ similar items to form $\mathcal{V}^{re}$, reducing the complexity to $\mathcal{O}(|\mathcal{V}|)$.

By incorporating approximation enhancement, the overall approximation objective in Eq. (\ref{eq:app}) can be rewritten as:

\begin{equation}
\label{eq:app_new}
\arg\min_{\widetilde{\mathbf{u}}_i^t, \widetilde{\mathbf{u}}_i^n} \{\mathcal{L}^{app}_i = \mathcal{L}^{app\_t}_i + \mathcal{L}^{app\_n}_i + \mathcal{L}_i^{dis}\},
\end{equation}

\subsubsection{Promotion enhancement}
Due to inherent item interaction overlap, perfectly separating real target and non-target groups during approximation is infeasible. As a consequence, conflicting optimization issues may arise in the promotion step. 
Concretely, the basic promotion step directly pulls the target group toward the target items while pushing the non-target group away. However, due to the entanglement of the two groups in the embedding space, these objectives may conflict, potentially pushing the target group away from the target items and resulting in a suboptimal solution.
To mitigate this conflict, we introduce enhancement strategies in both learning and optimization. The learning strategy aligns target and relevant item embeddings to strengthen their semantic connections, while the optimization strategy adaptively tunes coefficients to balance optimization directions across groups.

\noindent \textbf{Target\&Relevant item embedding alignment}. Motivated by popularity manipulation-based attacks \cite{DBLP:conf/www/YinXFG24}, which improve target item exposure by making them resemble popular items, we propose instead to enhance their exposure to the target user group by aligning them with relevant items. Specifically, we compute the cosine similarity between each target–relevant item pair and aim to maximize the average similarity across all pairs. Formally, the loss function is defined as follows:

\begin{equation}
\label{eq:similarity}
 \mathcal{L}^{sim}_i = \frac{1}{|\mathcal{V}^{re}||\widetilde{\mathcal{V}}|}\sum_{v_i \in \mathcal{V}^{re}}\sum_{v_j \in \widetilde{\mathcal{V}}} [1-\cos(\mathbf{v}_i, \mathbf{v}_j)].
\end{equation}
Minimizing Eq. (\ref{eq:similarity}) not only breaks the conflicting optimization dynamics but also improves the attack effectiveness on the target user group, as demonstrated in the experiments.

\noindent \textbf{Adaptive coefficient tuning.}
From the optimization perspective, a straightforward solution to mitigate the conflicting optimization is assigning different weights to different groups in Eq. (\ref{eq:pro}). For example, assigning a lower weight to the non-target group steers the optimization more toward associating target items with the target group. However, it is challenging for the attacker to determine appropriate weights, as they cannot validate the effectiveness of different configurations during federated training. Moreover, items with different frequencies exhibit varying update speeds \cite{DBLP:conf/icde/Zhang0RZ0S24}, and using a fixed weight may prevent the lower-weighted group from effectively building associations with items. 

Therefore, we propose adaptive coefficient tuning to adjust the optimization direction. The key idea is, if the target items are ranked as expected in a group's recommendation list (i.e., ranked high in the target group and low in the non-target group), then less optimization effort should be paid to that group. Based on this intuition, the weighting coefficients can be determined by the ranking performance of the target items within each group. Additionally, since the rankings of target items may be unreliable in the early stages of training, it is necessary to down-weight their influence during those rounds. Taking these  factors into account, we modify the promotion loss in Eq. (\ref{eq:pro}) as:
\begin{equation}
\label{eq:pro_adaptive}
\mathcal{L}^{pro\_ad}_i = (1+\gamma^{t}_e(\frac{e}{T})^2)\mathcal{L}^{pro\_t}_i+(1-\gamma^{n}_e(\frac{e}{T})^2)\mathcal{L}^{pro\_n}_i,
\end{equation}
where $T$ is the number of global epochs and $e$ is the current epoch. $\gamma^{t}_e$ and $\gamma^{n}_e$ is the target and non-target group ranking weights respectively. Specifically, $\gamma^{t}_e$ is defined as:
\begin{equation}
\label{eq:gamma}
\gamma^{t}_e = \frac{R^t_e}{R^t_e+R^n_e},
\end{equation}
where $R^t_e$ is the normalized average rank of target items in the approximated target users' recommender list and $R^n_e$ denotes that of the non-target users.

By combining promotion enhancement into the basic promotion step, the overall promotion objective in Eq. (\ref{eq:pro}) can be rewritten as:
\begin{equation}
\label{eq:pro_new}
\arg\min_{\widetilde{\mathbf{V}},\mathbf{\Theta}} \{\mathcal{L}^{pro}_i = \mathcal{L}^{pro\_ad}_i+\alpha \mathcal{L}^{sim}_i \},
\end{equation}
where $\alpha$ controls the weight of the target \& relevant item embedding alignment module.

\section{Experiments}
This section presents the primary experiment settings and results. Detailed settings and more experiments refer to Appendix.  

\subsection{Experiment settings}

\begin{table}[ht]                        
  \caption{Dataset Statistics}   
  \centering
  \renewcommand{\arraystretch}{1.2}   
  \begin{tabular}{lccccc}     
    \hline
    \textbf{Dataset} & \textbf{\# Users} & \textbf{\# Items} &
    \textbf{\# Interactions} & \textbf{\# Avg.} \\
    \hline
    ML-100K & 943   & 1,682 & 100,000  & 106 \\
    ML-1M   & 6,040 & 3,706 & 1,000,209 & 166 \\
    Steam   & 3,753 & 5,134 & 114,713   & 31 \\
    \hline
  \end{tabular}
    \label{tab:dataset}
\end{table}

\subsubsection{Datasets}
Following \cite{DBLP:conf/icde/RongYZYCH22}, we adopt three real-world datasets from two domains. The \textit{MovieLens-1M (ML-1M)} and \textit{MovieLens-100K (ML-100K)} \cite{DBLP:journals/tiis/HarperK16} are user-movie interactions, and \textit{Steam-200K (Steam)} \cite{DBLP:conf/www/CheuqueGP19} is user-game interactions. The statistics of the dataset are shown in Table \ref{tab:dataset}.

% We convert them into implicit data and use \textit{leave-one-out} to split training and test sets. 

% 

\subsubsection{Baselines}
We consider both attack and defense methods for FedRec as our baselines. For the attack methods, we consider \textit{FedRecAttack} \cite{DBLP:conf/icde/RongYZYCH22}, \textit{PipAttack} \cite{DBLP:conf/wsdm/ZhangYCHNC22}, \textit{A-hum} \cite{DBLP:conf/ijcai/RongHC22}, \textit{PSMU} \cite{DBLP:conf/sigir/YuanNHCY23}, \textit{PIECKUEA} \cite{DBLP:conf/icde/Zhang0RZ0S24}, \textit{PIECKIPE} \cite{DBLP:conf/icde/Zhang0RZ0S24} and \textit{PoisonFRS} \cite{DBLP:conf/www/YinXFG24}. We apply established defense methods \textit{NORMBOUND} \cite{DBLP:journals/corr/abs-1911-07963}, \textit{MEDIAN} \cite{DBLP:conf/icml/YinCRB18}, \textit{TRIMMEDMEAN} \cite{DBLP:conf/icml/YinCRB18}, \textit{KRUM} \cite{DBLP:conf/nips/BlanchardMGS17}, \textit{MULTIKRUM} \cite{DBLP:conf/nips/BlanchardMGS17}, \textit{BULYAN} \cite{DBLP:conf/icml/MhamdiGR18},and \textit{PIECK} \cite{DBLP:conf/icde/Zhang0RZ0S24}.

\renewcommand{\arraystretch}{1}
\begin{table*}[t]
\centering
\caption{Comparison of attack methods on attack performance (\%). Best results on target and all users are highlighted in bold.}
\begin{tabular}{lllcccccccc}
\hline
\multirow{2}{*}{\textbf{Dataset}} & \multirow{2}{*}{\textbf{Metric}} & \multirow{2}{*}{\textbf{Group}} & \multicolumn{8}{c}{\textbf{Attack Method}} \\
\cline{4-11}
 &  &  & PipAttack & FedRecAttack & A-hum & PSMU & PoisonFRS & PIECKUEA & PIECKIPE & Spattack \\
\hline
\multirow{6}{*}{ML-100K} 
& \multirow{3}{*}{ER@5}  & Target   & 100.00 & 0.00 & 100.00 & 100.00 & 0.00 & 100.00 & 0.00 & \textbf{100.00} \\
                          &                         & Non-target & 14.01 & 0.00 & 13.90 & 34.76 & 0.00 & 16.04 & 0.11 & 6.31 \\
                          &                         & All  & 93.00 & 50.00 & 93.05 & 82.62 & 50.00 & 91.98 & 49.95 & \textbf{96.85} \\
\cline{2-11}
& \multirow{3}{*}{ER@10} & Target   & 100.00 & 0.00 & 100.00 & 100.00 & 0.00 & 100.00 & 42.86 & \textbf{100.00} \\
                          &                         & Non-target & 73.15 & 0.00 & 72.94 & 97.43 & 0.00 & 79.46 & 0.96 & 38.40 \\
                          &                         & All  & 63.43 & 50.00 & 63.53 & 51.29 & 50.00 & 60.27 & 70.95 & \textbf{80.80} \\
\cline{1-11}
\multirow{6}{*}{ML-1M}   
& \multirow{3}{*}{ER@5}  & Target   & 100.00 & 0.00 & 100.00 & 100.00 & 10.11 & 100.00 & 0.00 & \textbf{100.00} \\
                          &                         & Non-target & 100.00 & 0.00 & 100.00 & 31.96 & 0.02 & 58.38 & 0.00 & 3.82 \\
                          &                         & All  & 50.00 & 50.00 & 50.00 & 84.02 & 55.05 & 70.81 & 50.00 & \textbf{98.09} \\
\cline{2-11}
& \multirow{3}{*}{ER@10} & Target   & 100.00 & 0.00 & 100.00 & 100.00 & 32.98 & 100.00 & 0.00 & \textbf{100.00} \\
                          &                         & Non-target & 100.00 & 0.00 & 100.00 & 88.12 & 0.07 & 100.00 & 0.00 & 12.28 \\
                          &                         & All  & 50.00 & 50.00 & 50.00 & 55.94 & 66.46 & 50.00 & 50.00 & \textbf{93.86} \\
\cline{1-11}
\multirow{6}{*}{Steam}   
& \multirow{3}{*}{ER@5}  & Target   & 0.00 & 0.00 & 100.00 & 1.63 & 0.00 & 27.87 & 0.00 & \textbf{100.00} \\
                          &                         & Non-target & 0.00 & 0.00 & 9.59 & 0.00 & 0.00 & 0.27 & 0.00 & 6.28 \\
                          &                         & All  & 50.00 & 50.00 & 95.21 & 50.82 & 50.00 & 63.80 & 50.00 & \textbf{96.86} \\
\cline{2-11}
& \multirow{3}{*}{ER@10} & Target   & 0.00 & 0.00 & 100.00 & 9.83 & 0.00 & 77.05 & 0.00 & \textbf{100.00} \\
                          &                         & Non-target & 0.00 & 0.00 & 100.00 & 0.05 & 0.00 & 4.09 & 0.00 & 57.72 \\
                          &                         & All  & 50.00 & 50.00 & 50.00 & 54.89 & 50.00 & \textbf{86.48} & 50.00 & 71.14 \\
\hline
\end{tabular}
\label{tab:attack effectiveness}
\end{table*}

\renewcommand{\arraystretch}{1}
\begin{table*}[h]
\centering
\caption{Comparison of attack methods on average training time per round (s).}
\begin{tabular}{ccccccccccc}
\hline
\textbf{Datasets}&  No-attack & PipAttack & FedRecAttack & A-hum & PSMU & PoisonFRS & PIECKUEA & PIECKIPE & Spattack \\
\hline
ML-100K & 1.0 & 1.4 & 2.9 & 1.1 & 1.6 & 1.2 & 0.9 & 1.3 & 1.1 \\
ML-1M  & 6.5 & 7.6 & 10.3 & 8.1 & 8.1 & 6.3 & 6.2 & 7.0 & 6.8 \\
Steam  & 4.0 & 3.4 & 7.0 & 5.6 & 4.9 & 3.5 & 3.6 & 3.6 & 4.6 \\
\hline
\end{tabular}
\label{fig:attack efficiency}
\end{table*}

\renewcommand{\arraystretch}{1}
\begin{table*}[t]
\centering
\caption{Comparison of attack methods on recommendation performance (\%).}
\resizebox{\textwidth}{!}{
\begin{tabular}{llccccccccc}
\hline
\textbf{Dataset} & \textbf{Metric} & No-attack &PipAttack & FedRecAttack & A-hum & PSMU & PoisonFRS & PIECKUEA & PIECKIPE & Spattack \\
\hline
\multirow{2}{*}{ML-100K} & HR@10 &8.484& 7.953 & 8.378 & 7.847 & 7.741 & 8.378 & 7.741 & 8.378 & 8.170 \\
                         & NDCG@10 &3.857& 3.702 & 3.825 & 3.665 & 3.578 & 3.825 & 3.627 & 3.825 & 3.770 \\
\hline
\multirow{2}{*}{ML-1M}   & HR@10 &7.831& 7.152 & 7.815 & 7.152 & 7.620 & 7.798 & 7.150 & 7.798 &7.848 \\
                         & NDCG@10 &4.097& 3.374 & 4.098 & 3.464 & 4.007 & 4.092 & 3.732 & 4.089 & 4.100 \\
\hline
\multirow{2}{*}{Steam}   & HR@10 &7.967& 7.780 & 7.967 & 6.395 & 7.914 & 7.967 & 7.967 & 7.967 & 7.860 \\
                         & NDCG@10 &3.593& 3.584 & 3.642 & 3.150 & 3.624 & 3.641 & 3.639 & 3.641 & 3.541 \\
\hline
\end{tabular}
}
\label{tab:recomendation performance}
\end{table*}

\subsubsection{Evaluation and implementations}
We evaluate the attack performance using ER@K defined in Eq. (\ref{eq:5}) and $\gamma$-GER@K defined in Eq.(\ref{eq:6}). For the recommendation performance, we employ HR@K and NDCG@K. The proportion of malicious users $\rho$ is set to 0.1\% for ML-1M. For the smaller ML-100K and Steam datasets, $\rho$ is 0.2\% to avoid having no malicious users.

\subsection{Attack effectiveness}
We evaluate the effectiveness of Spattack in terms of performance and efficiency, compared to existing attack methods. 
% More experiments can be found in Appendix \ref{sec:app:more_exp}.

\subsubsection{Attack performance against existing attack methods}
We evaluate the attack performance on different user groups under ER@5 and ER@10, and utilize 0.5-GER@K to evaluate the overall performance (\textit{All}). From Table \ref{tab:attack effectiveness} we observe: (1) Traditional full-group attacks have a poor attack trade-off between two groups. They generally exhibit a low 0.5-GER@K, particularly when $k=10$, indicating a lack of discrimination between the two groups. Noting that several attacks (e.g., PIECKUEA) achieve a higher 0.5-GER@K, but their limited effectiveness on the target group fails to satisfy real-world demands. (2) Spattack achieves consistently superior results on target groups, obtaining a 100\% ER@K across datasets, even under low malicious ratios (0.1\%). In contrast, several baselines are entirely ineffective on the target group (ER@K$=0$). This superiority mainly comes from our enhancement modules, as detailed in the ablation study. (3) Spattack presents a better trade-off between the two groups. It maintains high performance on the target group while significantly reducing the impact on the non-target group, better aligning with practical attack objectives.

\renewcommand{\arraystretch}{1}
\begin{table*}[t]
\centering
\caption{Comparison of Spattack against typical defense methods (\%).}
\begin{tabular}{lllcccccccc}
\hline
\multirow{2}{*}{\textbf{Dataset}} & \multirow{2}{*}{\textbf{Metric}} & \multirow{2}{*}{\textbf{Group}} & \multicolumn{7}{c}{\textbf{Defense Method}} \\
\cline{4-11}
 &  &  & Norm & Median & Trimmedmean & Krum & MultiKrum & Bulyan &PIECK&No-defense \\
\hline
\multirow{6}{*}{ML-100K} 
& \multirow{2}{*}{ER@5}  & Target   & 100.00 & 0.00 & 85.71 & 76.50 & 100.00 & 100.00 & 42.86 &100.00 \\
                          &                         & Non-target & 16.58 & 0.00 & 5.13 & 0.93 & 6.84 & 6.84 &1.07&6.31  \\
                          
\cline{2-11}
& \multirow{2}{*}{ER@10} & Target   & 100.00 & 0.00 & 100.00 & 100.00 & 100.00 & 100.00 & 71.43&100.00   \\
                          &                         & Non-target & 80.53 & 0.00 & 31.02 & 20.10 & 40.43 & 40.43 & 6.31&38.40 \\

\cline{2-11}
& \multirow{1}{*}{HR@10}  & All   & 7.85 & 6.89 & 8.17 & 8.06 & 8.06 & 8.06 & 7.74&8.17  \\  
% \cline{2-10}
& \multirow{1}{*}{NDCG@10}  & All   & 3.66 & 3.18 & 3.76 & 3.75 & 3.74 & 3.74 &3.42& 3.77 \\ 
\cline{1-11}
\multirow{6}{*}{ML-1M}   
& \multirow{2}{*}{ER@5}  & Target   & 60.44 & 0.00 & 100.00 & 100.00 & 89.74 & 76.92 & 61.54&100.00  \\
                          &                         & Non-target & 1.38 & 0.00 & 6.03 & 9.57 & 2.60 & 3.16 & 1.25&3.82  \\
                          
\cline{2-11}
& \multirow{2}{*}{ER@10} & Target   & 87.91 & 0.00 & 100.00 & 100.00 & 100.00 & 91.21 &100.00& 100.00 \\
                          &                         & Non-target & 5.50 & 0.00 & 16.00 & 24.5 & 8.35 & 10.93 &4.42& 12.28  \\
                          
\cline{2-11}
& \multirow{1}{*}{HR@10}  & All   & 7.78 & 6.85 & 7.80 & 7.50 & 7.81 & 7.76 & 6.92&7.85  \\
% \cline{2-10}
& \multirow{1}{*}{NDCG@10}  & All   & 4.08 & 3.11 & 4.09 & 3.95 & 4.08 & 4.06 &3.20& 4.10  \\  
\cline{1-11}
\multirow{6}{*}{Steam}   
& \multirow{2}{*}{ER@5}  & Target   & 32.79 & 100.00 & 100.00 & 100.00 & 10.00 & 91.21 & 50.82&100.00 \\
                          &                         & Non-target & 0.73 & 41.82 & 6.34 & 49.65 & 3.93 & 6.20 &3.63& 6.28  \\
                          
\cline{2-11}
& \multirow{2}{*}{ER@10} & Target   & 100.00 & 100.00 & 100.00 & 100.00 & 100.00 & 100.00 &100.00& 100.00  \\
                          &                         & Non-target & 10.97 & 100.00 & 58.53 & 100.00 & 29.71 & 100.00 & 25.19&57.72 \\
                          
\cline{2-11}
& \multirow{1}{*}{HR@10}  & All   & 6.50 & 6.21 & 7.43 & 6.95 & 7.78 & 7.70 & 7.62&7.86 \\
% \cline{2-10}
& \multirow{1}{*}{NDCG@10}  & All   &3.47 & 2.47 & 3.48 & 3.29 & 3.58 & 3.56 & 3.53&3.54  \\  
\hline
\end{tabular}
\label{tab:defense performance}
\end{table*}

\renewcommand{\arraystretch}{1}
\begin{table*}[t]
\centering
\caption{Ablation study of Spattack (\%).}
\begin{tabular}{lllccccccccc}
\hline
\multirow{2}{*}{\textbf{Dataset}} & \multirow{2}{*}{\textbf{Metric}} & \multirow{2}{*}{\textbf{Group}} & \multicolumn{7}{c}{\textbf{Variations of Spattack}} \\
\cline{4-11}
 &  &  & $origin$ &$-e$& $-appr\_e$ & $-promp\_e$ & $-appr\_e1$ & $-appr\_e2$ & $-promp\_e1$ & $-promp\_e2$   \\
\hline
\multirow{6}{*}{ML-1M}   
& \multirow{2}{*}{ER@5}  & Target   & 100.00 &0.00& 28.21 & 30.77 & 28.21 & 100.00 & 100.00 & 28.21 \\
                          &                         & Non-target & 3.82 &0.00& 0.07 & 0.17 & 0.05 & 4.82 & 5.58 & 0.08  \\                         
\cline{2-11}
& \multirow{2}{*}{ER@10} & Target   & 100.00 &0.00& 71.79 & 79.49 & 66.67 & 100.00 & 100.00 & 79.49 \\
                          &                         & Non-target & 12.28 &0.00& 0.80 & 1.22 & 0.52 & 14.07 & 15.10 & 1.00 \\
\cline{2-11}
& \multirow{1}{*}{HR@10}  & All   & 7.85 &7.85& 7.88 & 7.85 & 7.85 & 7.85 & 7.85 & 7.85  \\  
%\cline{2-10}
& \multirow{1}{*}{NDCG@10}  & All   & 4.10 &4.11& 4.12 & 4.11 & 4.10 & 4.10 & 4.10 & 4.11 \\
\cline{1-11}
\multirow{6}{*}{Steam}   
& \multirow{2}{*}{ER@5}  & Target   & 100.00 &0.00& 100.00 & 0.00 & 19.67 & 100.00 & 83.61 & 0.00 \\
                          &                         & Non-target & 6.28 &0.00& 9.59 & 0.00 & 0.08 & 8.15 & 4.98 & 0.00 \\
\cline{2-11}
& \multirow{2}{*}{ER@10} & Target   & 100.00 &0.00& 100.00 & 0.00 & 59.02 & 100.00 & 100.00 & 0.00 \\
                          &                         & Non-target & 57.72 &0.00& 100.00 & 0.00 & 2.63 & 78.39 & 41.25 & 0.00\\
\cline{2-11}
& \multirow{1}{*}{HR@10}  & All   & 7.86 &7.78& 6.37 & 7.97 & 8.01 & 6.37 & 7.78 & 7.91  \\  
%\cline{2-10}
& \multirow{1}{*}{NDCG@10}  & All   & 3.54 &3.58& 3.14 & 3.64 & 3.59 & 3.15 & 3.58 & 3.62 \\
\hline
\end{tabular}
\label{tab:ablaton study}
\end{table*}

\subsubsection{Attack efficiency against existing attack methods}
We conduct 30 rounds of training for all the baselines and report the average time per training round. The results are depicted in Table \ref{fig:attack efficiency}. Compared to vanilla FedRec (No-attack), Spattack introduces only a marginal overhead. Moreover, it achieves comparable or superior efficiency to other baselines. It is also observed that approximation-based methods (e.g., FedRecAttack and PSMU) require more training time than popularity-based methods, as approximating embeddings demands more local training rounds than directly aligning them. In contrast, Spattack significantly reduces training time by the proposed approximation enhancement. Overall, Spattack is practical and scalable for real-world deployment.

% DL-FRS requires more time compared to MF-FRS due to the usage of a learnable interaction function. Regarding the attack, we observe a negligible time increase for both MF-FRS and DL-FRS compared to the vanilla scenario (denoted as No(Att.& Def.)). This is because the three P IECK modules can be efficiently executed via matrix parallel. Fig. 6(b) also shows that P IECK U EA exhibits a bit higher time cost than P IECK I PE . This discrepancy arises from P IECK U EA requiring mining more popular items, and implementing multiple rounds in batches (default batch size is 5 and round size is 3) to improve the score of the target item on approximated user embeddings. For defense, the average time increase per round for MF-FRS and DL-FRS is only 0.47 and 1.13 seconds, respectively, compared to No(Att.& Def.). These slight increments are acceptable for maintaining FRS security. Overall, our attacks and defense are efficient and effective.

\subsection{Attack stealthiness}
We evaluate the stealthiness of Spattack by two ways \cite{DBLP:conf/icde/RongYZYCH22}: (1) recommendation performance under the attack, and (2) attack performance against defense methods.

\subsubsection{Recommendation performance} 
A significant degradation in recommendation performance can make the attack easily detectable by the server. Therefore, preserving recommendation performance is crucial for attack stealthiness. As shown in Table \ref{tab:recomendation performance}, all baselines suffer from either failing in attack effectiveness (e.g., PoisonFRS) or severely harming recommendation performance (e.g., PSMU), revealing a trade-off dilemma. In contrast, Spattack avoids this trade-off by focusing the attack on the target group while minimizing effects on others. As a result, Spattack still achieves competitive recommendation results on both datasets, and even achieves the best results on ML-1M, indicating the stealthiness of our attack.

\subsubsection{Attack performance against existing defense methods}
From Table \ref{tab:defense performance}, we can observe that Spattack demonstrates strong effectiveness, indicating that its gradients resemble benign ones rather than outliers, confirming its stealthiness. MEDIAN applies median aggregation, filtering out most updates and thereby weakening our attack. MEDIAN’s effect is amplified on denser datasets (ML-1M and ML-100K) and at very small malicious ratios where benign updates dominate. However, as the ratio increases ($\geq$ 1\%), MEDIAN becomes ineffective. Moreover, MEDIAN noticeably degrade recommendation performance, as it discards most updates, making it impractical for real-world deployment. Thus, our attack remains effective and stealthy in practice. Inspired by this observation, a potential mitigation strategy is to preserve the median update and selectively incorporate a few nearby updates, achieving a better trade-off between robustness and recommendation performance. Interestingly, some defenses, such as \textit{Bulyan}, even amplify the attack’s effect, as their indiscriminate gradient filtering and normalization weaken sparse benign updates, making malicious gradients dominate \cite{DBLP:conf/icde/Zhang0RZ0S24}.
%effect of sparse benign updates and makes malicious gradients dominate.

\begin{figure}[t]
    \centering
  \includegraphics[scale=.32]{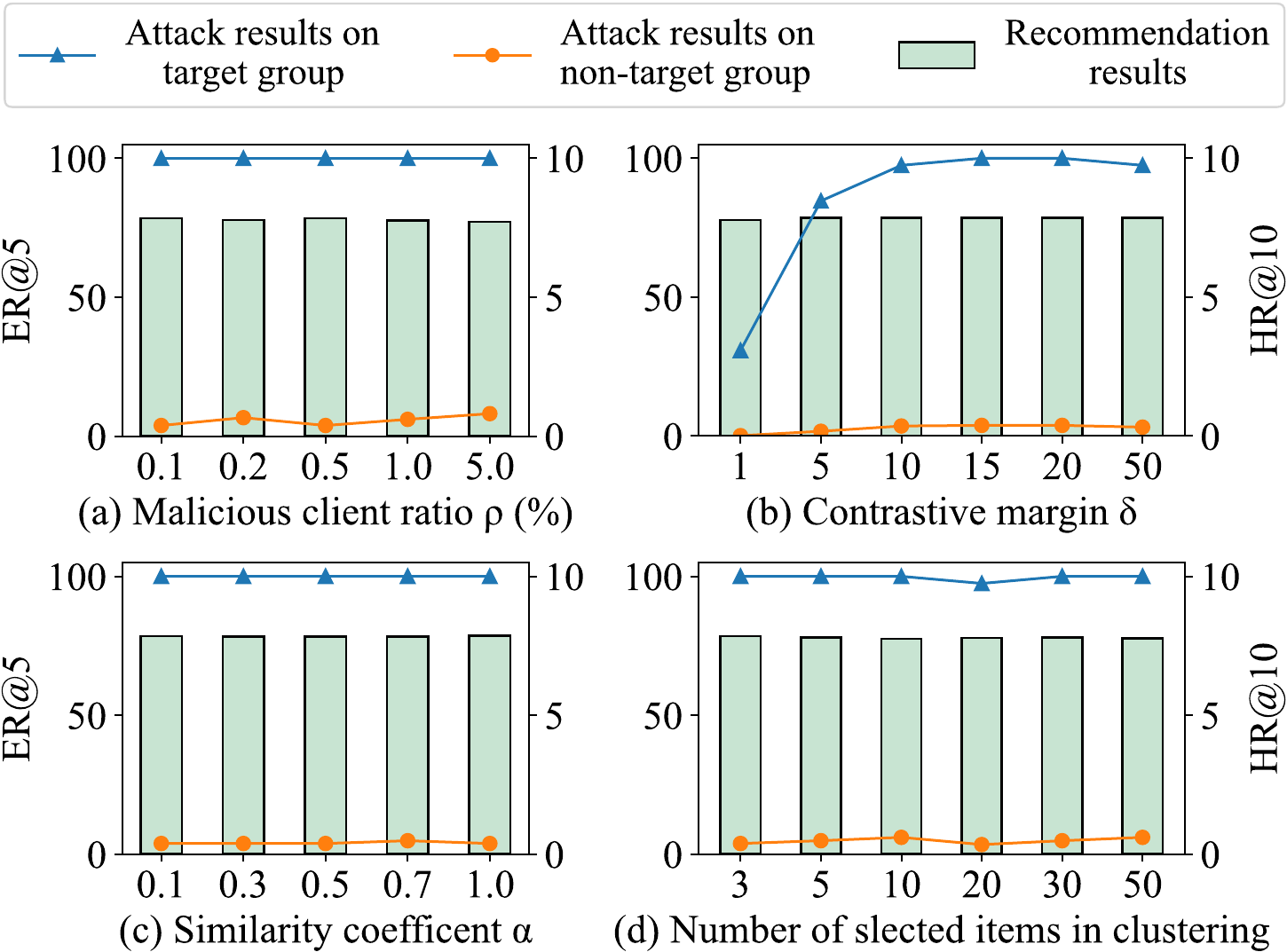}%[width=1\linewidth]
  \caption{Effects of key parameters in Spattack (\%).}
  \label{fig:parameter analysis}
\end{figure}

\subsection{Ablation study}
\label{sec:ablation study}
To evaluate the effects of key modules in Spattack, we design eight variations from five aspects: (1) The original Spattack ($origin$). (2) Removing all enhancements ($-e$). (3) Removing approximation enhancement ($-appr\_e$). (4) Removing promotion enhancement ($-promp\_e$). (5) Removing each sub-module of enhancement, including inter-group user embedding repulsion ($-appr\_e1$), relevant items construction ($-appr\_e2$), target \& relevant item embedding alignment ($-promp\_e1$), and adaptive coefficient tuning ($-promp\_e2$). From Table \ref{tab:ablaton study}, we draw the following conclusions.

(1) The basic approximate-and-promotion ($-e$) fails to achieve subgroup poisoning, results in complete ineffectiveness across all user groups. As discussed, the basic version lacks discriminative ability between the two groups, and also may cause conflicting optimization.

(2) $appr\_e$ improves the separation of target and non-target user groups. On ML-1M, removing it suffers significant attack performance decay on all groups, while on Steam, the performance improves. Without enhancement, entangled embeddings push both groups in the same direction and resulting in a poor trade-off. We also observe degraded recommendation performance on Steam, showing that it also harms model training.

(3) $promp\_e$ can steer the optimization of target items toward the desired direction. A dramatic drop in attack performance after removing the enhancement indicates that, without a reliable way to evaluate the approximated users, the basic promotion of target items may be misdirected. Noting that the recommendation performance slightly improves, as the optimization will focus more on the recommendation model.

(4) Each sub-module contributes to a distinct aspect of the enhancement. $appr\_e1$ improves attacks by directly pushing the two groups apart, thus may harm recommendations. $appr\_e2$ further clusters items to separate the groups semantically, reducing the negative impact. $promp\_e1$ aligns the target items with relevant items to guide the optimization. $promp\_e2$ further adaptively adjusts the direction, significantly improving the attack performance.

\subsection{Parameter analysis}

We analyze the key hyperparameters in Spattack on ML-1M, including the malicious client ratio $\rho$, contrastive margin $\delta$ in Eq. (\ref{eq:distance}), similarity coefficient $\alpha$ in Eq. (\ref{eq:pro_new}), and $k$ in top-$k$ most similar items in relevant items construction.  The results are shown in Fig. \ref{fig:parameter analysis}.

(1) \textbf{Malicious user ratios $\rho$}. 
From Fig. \ref{fig:parameter analysis}(a), we observe that increasing the malicious client ratio $\rho$ leads to stable attack performance on the target user group and a slight increase in the non-target group, while the recommendation performance remains unaffected, which demonstrates the strong robustness of Spattack.

(2) \textbf{Contrastive margin $\delta$}.
As shown in Fig.~\ref{fig:parameter analysis}(b), increasing $\delta$ initially improves attack performance on the target group, but excessive values lead to a slight decline. A small $\delta$ fails to sufficiently separate the two groups, while a large one may distort their semantic structure, harming performance.

(3) \textbf{Similarity coefficient $\alpha$}.
As shown in Fig. \ref{fig:parameter analysis}(c), both attack and recommendation performance are stable when increasing $\alpha$. Thus, $\alpha$ can be set arbitrarily as long as $\alpha > 0$. This flexibility is valuable in practice, as tuning hyperparameters is often challenging for attackers.

(4) \textbf{$k$ in relevant items construction}.
In Fig.~\ref{fig:parameter analysis}(d), the attack performance first decreases slightly and then increases. A small $k$ includes many items similar to the interested ones in the non-target approximation, while a large $k$ introduces dissimilar items into the target group approximation. Both reduce the distinction between groups, leading to an undesired performance increase.
\section{Conclusion}
In this paper, we present Spattack, the first poisoning attack that targets user subgroups in federated recommender systems. Following an approximate-and-promote paradigm, Spattack first separates target and non-target groups through contrastive repulsion and clustering-based item expansion, then aligns target and interested item embeddings while adaptively tuning the optimization weights across subgroups. In this way, a better trade-off between attack effectiveness against different user groups is achieved. Extensive experiments demonstrate that Spattack exhibits superior effectiveness and stealthiness. 
\begin{acks}
This work is supported in part by the National Natural
Science Foundation of China (No. 62550138, 62192784, 62572064, 62472329), JST PRESTO JPMJPR23P5, JST CREST JPMJCR21M2, and JST NEXUS JPMJNX25C4. This work is also supported in part by the Ministry of Education, Singapore, under its Academic Research Fund Tier 2 (Award MOE-T2EP20125-0005).
\end{acks}


\bibliographystyle{ACM-Reference-Format}
%\balance
\bibliography{sample-base}

\newpage
\appendix
%notations
\section{Notations}
\label{sec:app:notation}
Table~\ref{tab:notation} summarizes the frequently-used notations in our paper.

\begin{table}[h]
\begin{center}
\newcommand{\tabincell}[2]{\begin{tabular}{@{}#1@{}}#2\end{tabular}}
\caption{Summary of notations.}
\resizebox{1\linewidth}{!}{
\begin{tabular}{c|c}
\hline
\multicolumn{1}{c|}{\multirow{1}*{Notation}}&  \multicolumn{1}{c}{\multirow{1}*{Explanation}} \\
\hline
\multirow{1}*{}
$D_i$ & the private dataset for user (client) $u_i$ \\
$\mathcal{U}$, $\widetilde{\mathcal{U}}$  & the benign user set and malicious use set  \\
$\mathcal{U}^t$, $\mathcal{U}^n$  & the target user set and non-target use set \\

$\mathcal{V}$ & the item set \\
$\mathbf{V}$ & the item embedding table \\
$\Theta$ & the parameter of recommendation model\\
$\widetilde{\mathcal{V}}$ &  the target items \\
$\widetilde{\mathbf{V}}$ &  the target item embeddings table \\
$\mathcal{V}^{re}$  &  the relevant item set \\
$\mathcal{V}^{in}$  & the interested item set \\
$\widetilde{\mathbf{u}}^t$,$\widetilde{\mathbf{u}}^n$ & the approximated target/non-target user group embedding\\
$r_{ij}$, $\hat{r}_{ij}$ & the true rating and predicted rating of use $u_i$ to item $v_j$ \\
$\delta$ & the margin of contrastive-based distance \\
$\alpha$ & the weight coefficient of alignment loss \\
$T$ & the number of global epochs \\

\hline
\end{tabular}}
\label{tab:notation}
\end{center}
\end{table}

%algorithm
\section{Overall algorithm}
\label{sec:app:alg}
The overall algorithm of Spattack is presented in Algorithm \ref{alg:ours}. In each communication round $e$, the clients first receive the item embeddings $\mathbf{V}_e$ and global models $\mathbf{\Theta}_e$ from the server. Then, each malicious client locally performs approximation and promotion steps. At the approximation step, the client first constructs a relevant item set and then optimizes approximated user group embeddings  $\widetilde{\mathbf{u}}_i^t$ and $\widetilde{\mathbf{u}}_i^n$ using Eq. (\ref{eq:app_new}). At the promotion step, each client first calculates the ranks of target items based on $\mathbf{V}_e$ and $\mathbf{\Theta}_e$, then obtains the gradients $\mathbf{\nabla} \mathbf{\Theta}_e$ and $\mathbf{\nabla} \widetilde{\mathbf{V}}_e$ by optimizing Eq. (\ref{eq:pro_new}). Finally, the gradients are uploaded to the server for aggregation.

\begin{algorithm}[!ht]
  \renewcommand{\algorithmicrequire}{\textbf{Input:}}
  \renewcommand{\algorithmicensure}{\textbf{Output:}}
  \caption{Spattack} \label{alg:ours}
  \begin{algorithmic}[1]
    \Require global epoch $T$, local epoch $L$, interested item set $\mathcal{V}^{in}$, benign and malicious client sets $\mathcal{U}$, $\widetilde{\mathcal{U}}$
    \Ensure  server model $\mathbf{V}_T$, $\mathbf{\Theta}_{T}$
    \State Server initializes model parameters $\mathbf{V}_0$ and $\mathbf{\Theta}_0$
    \For {each round $e =0,\dots, T-1$}
      % \State if $t$ is the epoch when the attack starts, $\mathcal{U} = \mathcal{U} \cup \widetilde{\mathcal{U}}$
      \State Server distributes $\mathbf{V}_e$ and $\mathbf{\Theta}_e$ to $\widetilde{u}_i \in \widetilde{\mathcal{U}} \cup \mathcal{U}$.
        \For{$\widetilde{u}_i \in \widetilde{\mathcal{U}}$ \textbf{in parallel}} 
        \State // execute on client sides
        \State $\widetilde{\mathbf{u}}_i^t$, $\widetilde{\mathbf{u}}_i^n$ $\leftarrow$ \Call{Approximation}{$\mathbf{V}_e$, $\mathbf{\Theta}_e$}
        \State $\mathbf{\nabla} \widetilde{\mathbf{V}}_{i,e}$, $\mathbf{\nabla}\mathbf{\Theta}_{i,e}$ $\leftarrow$ \Call{Promotion}{$\widetilde{\mathbf{u}}_i^t$, $\widetilde{\mathbf{u}}_i^n$, $\mathbf{V}_e$, $\mathbf{\Theta}_e$, $e$}
        \EndFor
      % \State if $\widetilde{\mathcal{U}}_{t-1}$ is not empty, execute attack algorithm~\ref{alg_attack}
      \State // execute on central server
      \State Receive client gradients $\{\mathbf{\nabla}\mathbf{V}_{i,e}, \mathbf{\nabla} \mathbf{\Theta}_{i,e}\}_{u_{i}\in \widetilde{\mathcal{U}} \cup \mathcal{U}}$
      \State $\mathbf{V}_{e+1}$, $\mathbf{\Theta}_{e+1}$ $\leftarrow$ aggregation gradients
    \EndFor
  \Function{Approximation} {$\mathbf{V}$, $\mathbf{\Theta}$}
    \State Construct relevant item set $\mathcal{V}^{re}$ using $\mathbf{V}$ and $\mathcal{V}^{in}$
    \State Initialize user group embeddings $\widetilde{\mathbf{u}}_{0}^t$ and $\widetilde{\mathbf{u}}_{0}^n$
    \For{local round $j =0, \dots, L-1$} 
        \State $\widetilde{\mathbf{u}}_{j+1}^t$, $\widetilde{\mathbf{u}}_{j+1}^n$ $\leftarrow$ update embeddings by optimizing Eq. (\ref{eq:app_new})
        \EndFor
    \State \Return $\widetilde{\mathbf{u}}_L^t$, $\widetilde{\mathbf{u}}_L^n$
  \EndFunction
  \Function{Promotion} {$\widetilde{\mathbf{u}}^t$, $\widetilde{\mathbf{u}}^n$, $\mathbf{V}$, $\mathbf{\Theta}$, $e$}
    \State Calculate weights $\gamma^t_e$, $\gamma^n_e$ in Eq. (\ref{eq:gamma}) using $\widetilde{\mathbf{u}}^t$, $\widetilde{\mathbf{u}}^n$, $\mathbf{V}$, and $\mathbf{\Theta}$ 
    
    $\mathbf{\nabla} \widetilde{\mathbf{V}}$, $\mathbf{\nabla} \mathbf{\Theta}$ $\leftarrow$ obtain model gradients by optimizing Eq. (\ref{eq:pro_new})
    \State \Return $\mathbf{\nabla} \widetilde{\mathbf{V}}$, $\mathbf{\nabla} \mathbf{\Theta}$
  \EndFunction
    \end{algorithmic}
\end{algorithm}

%related works
\section{Related work}
\label{sec:related work}

\subsection{Recommender Systems}
Recommender systems have been extensively studied as a fundamental technique for modeling user preferences and delivering personalized recommendations \cite{Rendle2009BPR, DBLP:conf/sigir/Wang0WFC19, DBLP:journals/tkde/QiaoZLZBLW23, DBLP:conf/adma/BaoMYHS25}.
Traditional collaborative filtering methods~\cite{zou2020neural,chen2020efficient} rely on the explicit feedback, i.e., user-item ratings, and usually decompose the feedback into user embeddings and item embeddings by matrix factorization \cite{Koren2008Factorization, Rendle2009BPR}. With the development of deep learning, neural models such as NCF~\cite{DBLP:conf/www/HeLZNHC17} have been proposed to replace the linear interaction assumption of traditional matrix factorization with nonlinear neural architectures, enabling more expressive modeling of user–item relationships. However, these models often struggle to capture higher-order user–item relationships. To address this limitation, GNN-based methods model user–item interactions as a bipartite graph and propagate embedding information along its structure, effectively learning more expressive representations. For example, NGCF~\cite{DBLP:conf/sigir/Wang0WFC19} introduces explicit message passing to model collaborative signals between users and items, while LightGCN~\cite{DBLP:conf/sigir/0001DWLZ020} simplifies this process by removing nonlinear transformations to improve efficiency without sacrificing performance.

% Numerous data poisoning attacks have been proposed against centralized recommender systems \cite{DBLP:conf/icde/SongLHWLLG20, DBLP:conf/nips/LiWSV16, DBLP:conf/www/FangG020}.

\subsection{Attacks against Recommender Systems}
With the widespread adoption of recommender systems across various application domains, concerns regarding their security and robustness have received increasing attention \cite{li2016data, DBLP:conf/aaai/YanHYLCS25, DBLP:conf/kdd/Zhang0Z0Z0025, DBLP:conf/icde/SongLHWLLG20, DBLP:conf/nips/LiWSV16, DBLP:conf/www/FangG020}. Many works have demonstrated that recommender systems are susceptible to data poisoning attacks, which inject fake users with fabricated interactions to either promote target items (targeted attacks \cite{DBLP:conf/uss/FangCJG20, DBLP:conf/aaai/ZhangZWL00025}) or degrade overall recommendation performance (untargeted attacks \cite{DBLP:conf/wsdm/ZhangYCHNC22, DBLP:conf/icde/RongYZYCH22}). Targeted attack is typically more stealthy and harmful in practice, as it may go unnoticed while subtly biasing user experience.
Several studies have proposed data poisoning techniques targeting different classes of recommender systems~\cite{li2016data,huang2021data,fang2018poisoning}. However, these approaches generally exhibit limited effectiveness and require strong assumptions regarding the attacker’s prior knowledge. Specifically, works such as~\cite{li2016data, fang2018poisoning} assume that the attacker has access to the complete historical interaction data, while \cite{fang2020influence} requires access to a substantial portion of the interaction matrix. These assumptions are often unrealistic in practical deployment settings, where user interaction data is typically sparse, distributed, or privacy-protected.

\subsection{Federated recommendation}
Traditional recommendations hold a basic assumption that user interaction data can be collected centrally, which facilitates centralized model training \cite{DBLP:conf/www/HeLZNHC17, DBLP:conf/sigir/Wang0WFC19}. In reality, such setting largely violates the user privacy and even may be infeasible. To tackle this challenge, federated recommendation (FedRec) has emerged in recent years due to the property of collaborative training model without exposing user data \cite{DBLP:journals/corr/abs-1901-09888, DBLP:conf/www/YanCWYDS24, DBLP:conf/icde/QiaoZBZYW24}. Typically, each client trains the local model utilizing private user interaction data and only uploads the shared global model updates (e.g., item embeddings) to the server for aggregation, thus protecting privacy.

A spectrum of FedRec methods has been proposed to maintain recommendation performance while protecting user privacy \cite{DBLP:journals/expert/ChaiWCY21, DBLP:journals/corr/abs-1901-09888, DBLP:journals/expert/LinLPM21, DBLP:conf/aaai/LiangP021, DBLP:journals/corr/abs-2102-04925, DBLP:journals/tist/LiuYFPY22}. As a first work of FedRec based on the collaborative filter, FCF \cite{DBLP:journals/corr/abs-1901-09888} updates the user embeddings locally and uploads the gradients of item embeddings to the server for aggregation. FedMF \cite{DBLP:journals/expert/ChaiWCY21} further proves that only uploading item embedding gradients can also leak privacy, and thus encrypts the gradients with homomorphic encryption. To protect user interactions, FedRec \cite{DBLP:journals/expert/LinLPM21} proposes to upload gradients of randomly sampled items, and FedRec++ \cite{DBLP:conf/aaai/LiangP021} further utilizes denoising clients to eliminate the noise. Subsequently, some works utilize the graph to model the relation between users and items and apply GNNs to the FedRec \cite{DBLP:journals/corr/abs-2102-04925, DBLP:journals/tist/LiuYFPY22}, which achieves better recommendation results at the expense of computational efficiency and privacy guarantees.

\subsection{Attacks against federated recommendation}
Most centralized attacks rely on shared user interactions, which are not applicable to FedRec, as only gradients or model parameters can be shared. Nevertheless, recent studies have shown that FedRec remains vulnerable to gradient poisoning attacks, where attackers can upload malicious gradients to corrupt the model \cite{DBLP:conf/wsdm/ZhangYCHNC22, DBLP:conf/icde/RongYZYCH22}. Several targeted gradient poisoning attacks have been proposed to promote the exposure of target items in FedRec \cite{DBLP:conf/wsdm/ZhangYCHNC22, DBLP:conf/icde/RongYZYCH22, DBLP:conf/ijcai/RongHC22, DBLP:conf/sigir/YuanNHCY23, DBLP:conf/www/YinXFG24, DBLP:conf/icde/Zhang0RZ0S24, DBLP:journals/tifs/HaoCWLLWP24}. Among them, one line is based on the idea of promoting target items similar to popular items, the core of which is to find the popular items. To achieve this, PipAttack \cite{DBLP:conf/wsdm/ZhangYCHNC22} assumes that each item's popularity can be directly accessed by attackers, then the target item can be fed into a pretrained popularity classifier to maximize popularity. PoisonFRS \cite{DBLP:conf/www/YinXFG24} removes this prior knowledge by assuming that the average of all item embeddings is unpopular, and uses the distance to the averaged embedding as an indicator of item popularity. PIECK \cite{DBLP:conf/icde/Zhang0RZ0S24} further proposes to utilize the convergence speed and magnitude of gradients to identify popular items. Another line focuses on approximating benign users' embeddings and then promoting the target items to them. FedRecAttack \cite{DBLP:conf/icde/RongYZYCH22} leverages some public interactions to approximate user embeddings. To eliminate the requirement for public interactions, A-hum \cite{DBLP:conf/ijcai/RongHC22} and PSMU \cite{DBLP:conf/sigir/YuanNHCY23} approximate user embeddings through random initialization or randomly selected interaction items. Unlike existing attacks that promote target items to all users, this work focuses on targeting a specific user subgroup, which better aligns with real-world scenarios, reduces the impact on recommendation performance, and enhances attack stealth.

%datasets
\section{Detailed experiment settings}
\label{sec:app:settings}

\renewcommand{\arraystretch}{1}
\begin{table*}[t]
\centering
\caption{Comparison of attack methods under different $\gamma$ ($\gamma$-GER@5) (\%). Best results are highlight in bold.}
\begin{tabular}{cccccccccc}
\hline
$\gamma$& PipAttack & FedRecAttack & A-hum & PSMU & PoisonFRS & PIECKUEA & PIECKIPE & Spattack \\
\hline
0 & 85.99 & \textbf{100.00} & 86.10 & 65.24 & \textbf{100.00} & 83.96 & 99.89 & 93.69 \\
0.3 & 90.19 & 70.00 & 90.27 & 75.67 & 70.00 & 88.77 & 69.92 & \textbf{95.58} \\
0.5  & 93.00 & 50.00 & 93.05 & 82.62 & 50.00 & 91.98 & 49.95 & \textbf{96.85} \\
0.7  & 95.797 & 30.00 & 95.83 & 89.57 & 30.00 & 95.19 & 29.97 & \textbf{98.11} \\
1  & 100.00 & 0.00 & 100.00 & 100.00 & 0.00 & 100.00 & 0.00 & \textbf{100.00} \\
\hline
\end{tabular}
\label{tab:app:gamma}
\end{table*}

\subsection{Baselines}
The brief introductions of the baselines are as follows.

\noindent \textbf{FedRecAttack} \cite{DBLP:conf/icde/RongYZYCH22}. It uses public data to estimate benign users and optimizes objectives that increases the targeted item's popularity.

\noindent \textbf{PipAttack} \cite{DBLP:conf/wsdm/ZhangYCHNC22}. It leverages item popularity to build a popularity estimator and generates model updates pushing the target item toward higher popularity.

\noindent \textbf{A-hum} \cite{DBLP:conf/ijcai/RongHC22}. It randomly initializes embeddings for malicious users and enhances the attack by mining hard-to-influence users.

\noindent \textbf{PSMU} \cite{DBLP:conf/sigir/YuanNHCY23}. It synthesizes per-round local data for fake users, uses their features to measure item popularity. 

\noindent \textbf{PIECKIPE} \cite{DBLP:conf/icde/Zhang0RZ0S24}. It promotes target items by aligning their embeddings with those of mined popular items using a similarity loss.

\noindent \textbf{PIECKUEA} \cite{DBLP:conf/icde/Zhang0RZ0S24}. It approximates user embeddings with popular item embeddings to directly raise target item scores, achieving stronger performance than \textsc{PIECKIPE}.

\noindent \textbf{PoisonFRS} \cite{DBLP:conf/www/YinXFG24}. It uses item embeddings to build a scaled target embedding from estimated popular items and repeatedly uploads crafted updates with filler interactions to promote target items.

\noindent \textbf{NORMBOUND} \cite{DBLP:journals/corr/abs-1911-07963}. The $L_2$ norm of each user’s uploaded gradient is restricted by a predefined threshold.

\noindent \textbf{MEDIAN} \cite{DBLP:conf/icml/YinCRB18}. The median value of the received gradients is computed for each dimension.

\noindent \textbf{TRIMMEDMEAN} \cite{DBLP:conf/icml/YinCRB18}. The $\tilde{p}$ largest and smallest values in each dimension are removed, the remains are averaged.

\noindent \textbf{KRUM} \cite{DBLP:conf/nips/BlanchardMGS17}. Among all gradients, the one with the smallest squared Euclidean distance to the others is selected.

\noindent \textbf{MULTIKRUM} \cite{DBLP:conf/nips/BlanchardMGS17}. The $2\tilde{p}$ least similar gradients identified by \textsc{KRUM} are iteratively removed, the remains are averaged.

\noindent \textbf{BULYAN} \cite{DBLP:conf/icml/MhamdiGR18}. Gradients are first selected by \textsc{MultiKrum} and then aggregated using \textsc{TrimmedMean}.

\noindent \textbf{PIECK} \cite{DBLP:conf/icde/Zhang0RZ0S24}. It is designed for FedRecs, introducing regularizations to confuse popular and unpopular items and separate user and popular items to reduce popularity bias.

%implement details
\subsection{Implementation Details}
\label{sec:app:implementation}
We convert all the datasets into implicit data and use \textit{leave-one-out} to split training and test sets. To select interested items, To select interested items, we avoid low-occurrence-frequency items to ensure a sufficient number of target users. Specifically, we compute the frequency of all items and randomly sample $m$ items with frequencies in the range [0.2,1] as interested items. We set $m=10$ for ML-100K and ML-1M, and $m=5$ for Steam to maintain an appropriate number of interested items. The number of approximate users in each group is 10.
For all baselines, the dimension of user and item embedding is 8. The batch size is 256, and the learning rate is 0.001. The number of global epochs is 30 to ensure convergence. The ratio of positive to negative samples for each user is 1:1. Unless otherwise specified, we adopt a two-layer MLP with a hidden size of 8 as the recommendation model.  Following \cite{DBLP:conf/icde/RongYZYCH22}, all the results are evaluated at the last training epoch.

\section{More Experimental Results}
\label{sec:app:more_exp}

\subsection{Performance under different $\gamma$ in $\gamma$-GER@5}

\begin{table}[t]
\centering
\caption{Effects of different number of interested items (\%). }
\resizebox{\columnwidth}{!}{
\begin{tabular}{lllccc}
\toprule
\multicolumn{3}{l}{\textbf{\# Interested items (\# target users)}} & \textbf{5 (30)} & \textbf{10 (7)} & \textbf{15 (2)} \\
\midrule
\multirow{4}{*}{\textbf{Attack}} 
  & \multirow{2}{*}{Target}   & ER@5    & 100.00 & 100.00 & 100.00 \\
  &                              & ER@10   & 100.00 & 100.00 & 100.00 \\
\cline{2-6}
  & \multirow{2}{*}{Non-target} & ER@5    & 10.23  & 6.31   & 5.74   \\
  &                              & ER@10   & 47.21  & 38.40  & 31.70  \\
\midrule
\multirow{2}{*}{\textbf{Rec}} 
  & \multirow{2}{*}{All}     & HR@10   & 8.17   & 8.17   & 8.16   \\
  &                              & NDCG@10 & 3.77   & 3.77   & 3.72   \\
\bottomrule
\end{tabular}
}
\label{tab:app:interested}
\end{table}

In our proposed metric $\gamma$-GER@5, the parameter $\gamma \in [0, 1]$
can be adjusted to accommodate different practical requirements. A smaller $\gamma$ places greater emphasis on attack stealthiness on the non-targeted group, while a larger one focuses more on attack performance on the targeted group. To evaluate the effectiveness of Spattack under different requirements, we vary $\gamma$ and compare the results with the baselines on the ML-100K dataset. As shown in Table \ref{tab:app:gamma}, Spattack achieves relatively stable results across various $\gamma$, which enables Spattack to satisfy diverse practical requirements under different application scenarios.

\subsection{Performance under different number of interested items}

\begin{table}[t]
\centering
\caption{Comparison of attack methods on GNN-based FedRecs (\%). Best results on all users are highlight in bold.}
\resizebox{\columnwidth}{!}{
\begin{tabular}{lllccc}
\toprule
& \textbf{Group}& \textbf{Metric} & A-hum & PIECKUEA & Spattack \\
\midrule
\multirow{6}{*}{\textbf{Attack}} 
  & \multirow{2}{*}{Target}   & ER@5    & 100.00 & 100.00 & 95.32 \\
  &                              & ER@10   & 100.00 & 100.00 & 100.00 \\
  \cline{2-6}
  & \multirow{2}{*}{Non-target} & ER@5    & 30.91  & 25.17   & 6.89   \\
  &                              & ER@10   & 89.23  & 81.56  & 41.56  \\
  \cline{2-6}
  & \multirow{2}{*}{All} & 0.5-GER@5    & 84.55  & 87.42   & \textbf{94.22}   \\
  &                              & 0.5-GER@10   &55.39& 59.22  & \textbf{79.22} 
  \\
\midrule
\multirow{2}{*}{\textbf{Rec}} 
  & \multirow{2}{*}{All}     & HR@10   & 7.99   & 7.89   & 7.98   \\
  &                              & NDCG@10 & 3.56   & 3.58   & 3.55   \\
\bottomrule
\end{tabular}
}
\label{tab:app:gnn}
\end{table}

In practice, the attack may choose different number of interested items to identify the target user groups. To show the effects, We randomly sample different numbers of interested items and conduct experiments on the ML-100K dataset. Table \ref{tab:app:interested} shows that increasing the number of interested items reduces the attack performance on the non-target group, while the target group performance and recommendation quality remain stable. According to the definition of the target user group, a larger number of interested items results in a smaller target group and consequently a larger non-target group, which may dilute the attack effect on the latter. Moreover, increasing the number of interested items allows for more precise identification of target users. Overall, Our attack remains robust across different numbers of interested items, enabling attackers to flexibly choose the quantity in practical scenarios.

%visualization
\begin{figure}[t]
    \centering
  \includegraphics[scale=.27]{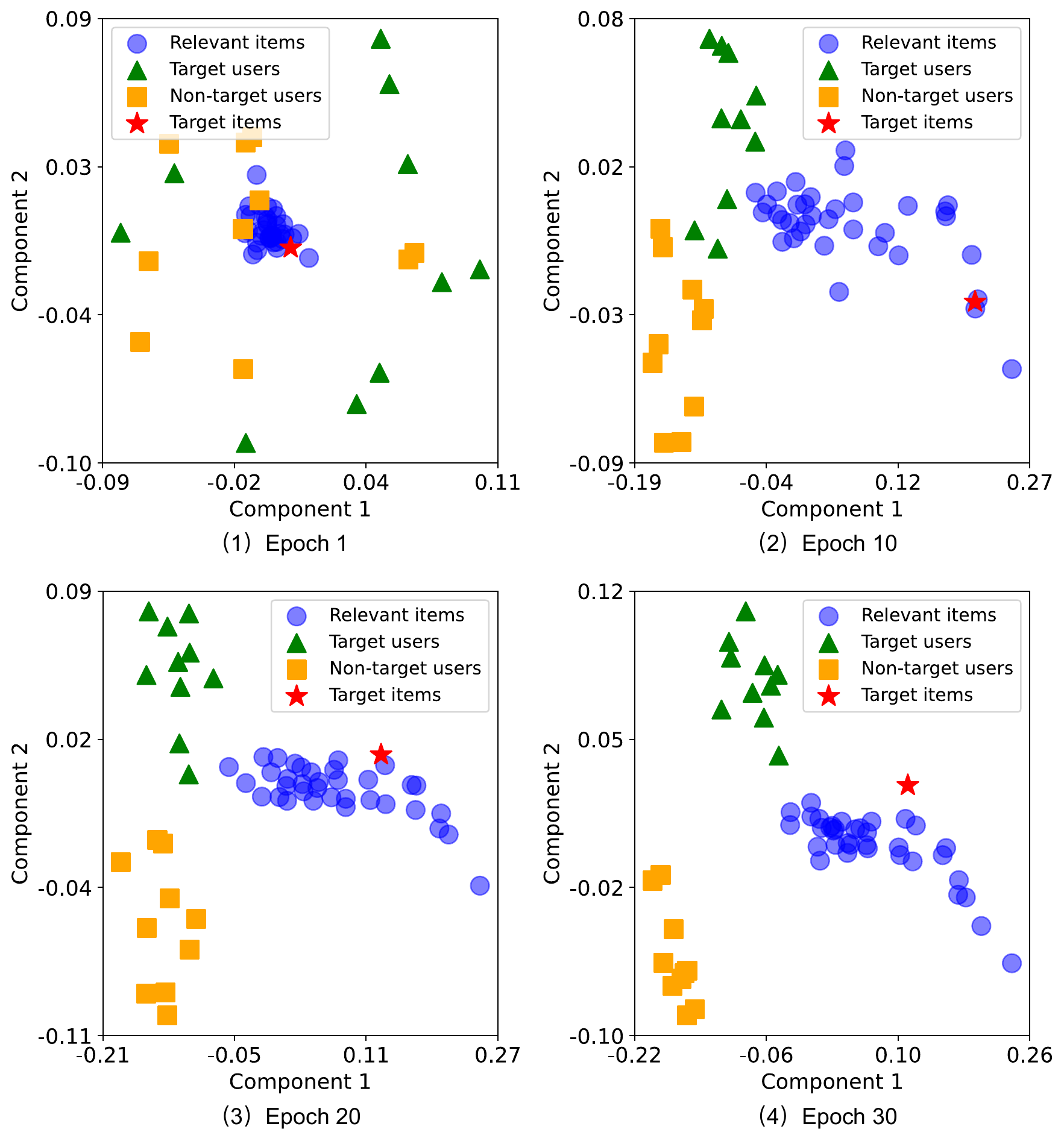}%[width=1\linewidth]
  \caption{Visualization of embeddings during training.}
  \label{fig:vis}
\vspace{-5mm}
\end{figure}

\subsection{Performance on GNN-based FedRecs}
To evaluate the effectiveness of Spattack on other FedRec architectures, we adopt the typical GNN-based FedRec framework \cite{DBLP:journals/corr/abs-2102-04925} and conduct experiments on the ML-100K dataset. We compare Spattack with two baselines A-hum \cite{DBLP:conf/ijcai/RongHC22} and PIECKUSE \cite{DBLP:conf/icde/Zhang0RZ0S24}, which represent typical approximation-based and popularity-based methods respectively. As shown in Table~\ref{tab:app:gnn}, Spattack consistently outperforms both baselines on 0.5-GER@5 and 0.5-GER@10, and achieves comparable recommendation results, demonstrating its adaptability across different FedRec architectures.

\subsection{Visualization}
\label{sec:visualization}
To provide a clearer understanding of the underlying mechanism of Spattack, we visualize the embeddings from 1, 10, 20, and 30 epochs on Steam using t-SNE, as shown in Fig. \ref{fig:vis}. 
Initially, the user embeddings of the two groups are highly entangled. The embeddings gradually diverge due to the inter-group user embeddings repulsion. The clustering strategy ensures that relevant items stay close in each round, while the target–relevant item alignment further pulls target items toward them. During training, adaptive coefficient tuning guides the optimization to better balance the two groups, gradually pulling target items closer to the target user group.

\end{document}